\documentclass[aps,pra,10pt,nofootinbib]{revtex4-2}
\usepackage{amsfonts,amssymb,amsmath}            
\usepackage{lmodern,adjustbox}
\usepackage[skins,breakable]{tcolorbox}
\usepackage{tikz}
\usepackage{enumitem}
\usepackage{fontawesome}
\usepackage{marginnote}
\usetikzlibrary{quantikz2}
\tcbuselibrary{listings}
\tcbuselibrary{breakable}
\usetikzlibrary{shapes.geometric}
\usepackage[pdftex,hyperfigures,pdfpagelabels]{hyperref} 

\usetikzlibrary{external}

\NewTCBListing{Code}{ !O{0.4} }{enhanced,fonttitle=\sffamily\bfseries\large,valign=center,drop fuzzy shadow,lefthand ratio=#1,lower separated=false,text side listing,before upper=\centering}
\newtcolorbox{FullCode}{enhanced,fonttitle=\sffamily\bfseries\large,valign=center,drop fuzzy shadow,breakable}
\tcbset{shield externalize} 

\makeatletter

\pgfaddtoshape{isosceles triangle}{
  \anchor{lstartone}{%
    \trianglepoints 
    \pgf@process{\pgfpointadd{\centerpoint}{\pgfmathrotatepointaround{\lowerleftanchor}{\pgfpointorigin}{\rotate}}}
    \pgf@ya=.5\pgf@y
    \pgf@process{\pgfpointadd{\centerpoint}{\pgfmathrotatepointaround{\lowerrightanchor}{\pgfpointorigin}{\rotate}}}
    \pgf@y=.5\pgf@y
    \advance\pgf@y by \pgf@ya%
    \advance\pgf@y by -\pgfkeysvalueof{/tikz/commutative diagrams/classical gap vertical}%
  }
  \anchor{lstarttwo}{%
    \trianglepoints
    \pgf@process{\pgfpointadd{\centerpoint}{\pgfmathrotatepointaround{\lowerleftanchor}{\pgfpointorigin}{\rotate}}}
    \pgf@ya=.5\pgf@y
    \pgf@process{\pgfpointadd{\centerpoint}{\pgfmathrotatepointaround{\lowerrightanchor}{\pgfpointorigin}{\rotate}}}
    \pgf@y=.5\pgf@y
    \advance\pgf@y by \pgf@ya%
    \advance\pgf@y by \pgfkeysvalueof{/tikz/commutative diagrams/classical gap vertical}%
  }
  \anchor{lendone}{%
    \trianglepoints
    \pgf@process{\pgfpointintersectionoflines%
      {\pgfpointadd{\centerpoint}{\pgfmathrotatepointaround{\apexanchor}{\pgfpointorigin}{\rotate}}}%
      {\pgfpointadd{\centerpoint}{\pgfmathrotatepointaround{\lowerrightanchor}{\pgfpointorigin}{\rotate}}}%
      {\pgfpointadd{\centerpoint}{\pgfpoint{0cm}{-\pgfkeysvalueof{/tikz/commutative diagrams/classical gap vertical}}}}%
      {\pgfpointadd{\centerpoint}{\pgfpoint{0.5cm}{-\pgfkeysvalueof{/tikz/commutative diagrams/classical gap vertical}}}}%
    }%
  }
  \anchor{lendtwo}{%
    \trianglepoints
    \pgf@process{\pgfpointintersectionoflines%
      {\pgfpointadd{\centerpoint}{\pgfmathrotatepointaround{\apexanchor}{\pgfpointorigin}{\rotate}}}%
      {\pgfpointadd{\centerpoint}{\pgfmathrotatepointaround{\lowerleftanchor}{\pgfpointorigin}{\rotate}}}%
      {\pgfpointadd{\centerpoint}{\pgfpoint{0cm}{\pgfkeysvalueof{/tikz/commutative diagrams/classical gap vertical}}}}%
      {\pgfpointadd{\centerpoint}{\pgfpoint{0.5cm}{\pgfkeysvalueof{/tikz/commutative diagrams/classical gap vertical}}}}%
    }%
  }
  \anchor{dstarttwo}{%
    \trianglepoints
    \pgf@process{\pgfpointintersectionoflines%
      {\pgfpointadd{\centerpoint}{\pgfmathrotatepointaround{\apexanchor}{\pgfpointorigin}{\rotate}}}%
      {\pgfpointadd{\centerpoint}{\pgfmathrotatepointaround{\lowerrightanchor}{\pgfpointorigin}{\rotate}}}%
      {\pgfpointadd{\centerpoint}{\pgfpoint{\pgfkeysvalueof{/tikz/commutative diagrams/classical gap horizontal}}{0cm}}}%
      {\pgfpointadd{\centerpoint}{\pgfpoint{\pgfkeysvalueof{/tikz/commutative diagrams/classical gap horizontal}}{0.5cm}}}%
    }%
  }
  \anchor{dstartone}{%
    \trianglepoints
    \pgf@process{\pgfpointintersectionoflines%
      {\pgfpointadd{\centerpoint}{\pgfmathrotatepointaround{\apexanchor}{\pgfpointorigin}{\rotate}}}%
      {\pgfpointadd{\centerpoint}{\pgfmathrotatepointaround{\lowerrightanchor}{\pgfpointorigin}{\rotate}}}%
      {\pgfpointadd{\centerpoint}{\pgfpoint{-\pgfkeysvalueof{/tikz/commutative diagrams/classical gap horizontal}}{0cm}}}%
      {\pgfpointadd{\centerpoint}{\pgfpoint{-\pgfkeysvalueof{/tikz/commutative diagrams/classical gap horizontal}}{0.5cm}}}%
    }%
  }
    \anchor{dendtwo}{%
    \trianglepoints
    \pgf@process{\pgfpointintersectionoflines%
      {\pgfpointadd{\centerpoint}{\pgfmathrotatepointaround{\apexanchor}{\pgfpointorigin}{\rotate}}}%
      {\pgfpointadd{\centerpoint}{\pgfmathrotatepointaround{\lowerleftanchor}{\pgfpointorigin}{\rotate}}}%
      {\pgfpointadd{\centerpoint}{\pgfpoint{\pgfkeysvalueof{/tikz/commutative diagrams/classical gap horizontal}}{0cm}}}%
      {\pgfpointadd{\centerpoint}{\pgfpoint{\pgfkeysvalueof{/tikz/commutative diagrams/classical gap horizontal}}{0.5cm}}}%
    }%
  }
  \anchor{dendone}{%
    \trianglepoints
    \pgf@process{\pgfpointintersectionoflines%
      {\pgfpointadd{\centerpoint}{\pgfmathrotatepointaround{\apexanchor}{\pgfpointorigin}{\rotate}}}%
      {\pgfpointadd{\centerpoint}{\pgfmathrotatepointaround{\lowerleftanchor}{\pgfpointorigin}{\rotate}}}%
      {\pgfpointadd{\centerpoint}{\pgfpoint{-\pgfkeysvalueof{/tikz/commutative diagrams/classical gap horizontal}}{0cm}}}%
      {\pgfpointadd{\centerpoint}{\pgfpoint{-\pgfkeysvalueof{/tikz/commutative diagrams/classical gap horizontal}}{0.5cm}}}%
    }%
  }
}

\pgfdeclareanchoralias{isosceles triangle}{dendtwo}{ustartwo}
\pgfdeclareanchoralias{isosceles triangle}{dendone}{ustartone}
\pgfdeclareanchoralias{isosceles triangle}{dstarttwo}{uendtwo}
\pgfdeclareanchoralias{isosceles triangle}{dstartone}{uendone}
\pgfdeclareanchoralias{isosceles triangle}{lendtwo}{rstartwo}
\pgfdeclareanchoralias{isosceles triangle}{lendone}{rstartone}
\pgfdeclareanchoralias{isosceles triangle}{lstarttwo}{rendtwo}
\pgfdeclareanchoralias{isosceles triangle}{lstartone}{rendone}
\makeatother

\tikzset{
  trimeter/.style={
      thickness, 
      filling,
      shape=isosceles triangle, 
      draw=black, 
      inner sep=2pt
  }}
\DeclareExpandableDocumentCommand{\mygate}{O{}{m}}{|[trimeter,#1]| {#2}}

\begin{document}

\title{Tutorial on the Quantikz Package}
\date{\today}
\author{Alastair \surname{Kay}}
\affiliation{Royal Holloway University of London, Egham, Surrey, TW20 0EX, UK}
\email{alastair.kay@rhul.ac.uk}
\begin{abstract}
\end{abstract}
\maketitle
I've always used \href{https://arxiv.org/abs/quant-ph/0406003}{QCircuit} for typesetting quantum circuit diagrams within \LaTeX, but found the Xy-pic based notation rather impenetrable and I struggled to adapt it for my needs (this is probably my failing rather than the package's). Thus, I wanted a tikz package that could do the same. That package is Quantikz. Those familiar with QCircuit will recognise much of the notation, although it has evolved a bit (hopefully simplified!).

The latest release (denoted by version numbers 1.x) is a major step forward in terms of the behind-the-scenes code. Unfortunately, this has necessitated breaking some compatibility with previous versions. Your old circuits should still work, but they might not look exactly as expected! Primarily the concept of the wires in a circuit has been modified as classical wires were only ever an after-thought, but have now been elevated to an equivalent status to the quantum wire.

\tableofcontents

\section{Usage}

The quantikz package is available on CTAN, and will therefore be available through most (current) TeX distributions. Once installed, simply write
\begin{verbatim}
\usepackage{tikz}
\usetikzlibrary{quantikz2}
\end{verbatim}
in the preamble of your document.
Now, each time that you want to include a quantum circuit, you just enclose it in a \verb!quantikz! environment. 

When new versions of quantikz are released, there can be a lag before the new features are reflected in the version used on the arXiv. If you need to provide the latest version, then simply include the file tikzlibraryquantikz2.code.tex (you will always be able to locate it on your computer in the main tex directory if you have installed the package, but it should also accompany the source code of this file, and the most recent version is available \href{https://ctan.org/tex-archive/graphics/pgf/contrib/quantikz}{here}) in the main directory of your source code.

\section{Basic Usage}

The simplest way for a new user to get started may be through the \href{https://www.ma.rhul.ac.uk/akay/quantikz/}{web-based editor} -- drag and drop some of the basic circuit elements, and it will automatically output for you a piece of code to paste into your tex document. More details can be found in Sec.\ \ref{sec:web}.

Quantum circuits are laid out with a matrix notation, with cells separated by \& (just like all matrices, tables etc.\ in \LaTeX). Here, we typeset a single quantum wire.
\begin{Code}
\begin{quantikz}
&&&&&
\end{quantikz}
\end{Code}

New wires are created by the new line command, \textbackslash\textbackslash. By default, all wires are quantum wires, i.e.\ a single solid line.
\begin{Code}
\begin{quantikz}
&&&&& \\
&&&&&
\end{quantikz}
\end{Code}
Wire types can be changed at a global level. Options are \texttt{q} (quantum), \texttt{c} (classical), \texttt{b} (bundle) and \texttt{n} (none). Here we specify that the first wire is quantum, and the second is classical. We also show how to change the classical wire to a quantum one mid-circuit (without separating gates, it looks ugly).
\begin{Code}
\begin{quantikz}[wire types={q,c}]
&&&&& \\
&&&&\setwiretype{q}&
\end{quantikz}
\end{Code}
The wire change command affects the wire drawn from that cell, which is the one running back to the previous cell. This may mean changing the wire type in the column \emph{after} the one you're expecting!

\subsection{Gates \& Measurement}

Inside a cell (between two \&), you can insert a gate command. Often, this will just be \verb|\gate|, a plain box that contains a label corresponding to the parameter. The first optional parameter can be used to specify the number of qubits that the gate spans. You always put the command in the first cell in which that gate should appear. The label of the gate command is already in math mode, so you can enter arbitrary mathematical functions.
\begin{Code}
\begin{quantikz}
&\gate{H}&\gate[2]{U}&\gate{R_Z(\theta)}& \meter{} \\
&&&\phase{\alpha} &
\end{quantikz}
\end{Code}
You can also see in the above example how to add a measurement gate (of which there are a number of variants) and a phase gate.

It is important that gate commands come before any other commands in a cell, such as those changing wire type. You do not typically put a gate command in the first cell (the one before the first \& in a row) as there won't be an entering wire in that case.

Make sure you include the trailing \& after the gate, or you won't have a wire coming out of the last gate!

\subsection{Controlled Gates}

Controlled gates typically consist of two elements -- a control and a target. The control creates a vertical wire of a specified length (may be negative). Note that the target commands, although they don't accept a parameter, need the \{\} after their call. Standard gates can also be targets.
\begin{Code}
\begin{quantikz}
& \ctrl{1} & \targ{} & \swap{1} & \ctrl[vertical wire=c]{2} &&\\
& \control{} & \ctrl[open]{-1} & \targX{} && \gate{X} &\\
&&&& \gate{U} & \meter{} \wire[u][1]{c}
\end{quantikz}
\end{Code}
Other controls can be created simply by the addition of a vertical wire.
If you want a gate that is one control and several targets, it is often good enough to just create the vertical wire so that it goes to the most distant target. You may need the command \verb!\wire[d]{q}! (vertical quantum wire) to create vertical connections.

\begin{Code}
\begin{quantikz}
& \ctrl{2} & \gate{U} & \\
& \targ{} & \ctrl{1}\wire[u]{q} & \\
& \targ{} & \gate{V} &
\end{quantikz}
\end{Code}


\subsection{Labelling Circuits}

There are many ways to label the different parts of your circuit. The basic commands for labelling wires at the start/end of a circuit are \verb!\lstick! and \verb!\rstick! respectively. These can apply to multiple qubits; just apply the command to the first, and say how many wires it should cover. The arguments are typset in text mode, but you can of course convert to math mode.
\begin{Code}
\begin{quantikz}
\lstick{\ket{0}} & \gate{U} & \gate[2]{\sqrt{\textsc{swap}}} &\rstick[2]{out} \\
\lstick[2]{input} & \gate[2]{V} && \\
& & & \rstick{$\frac{1}{\sqrt{2}}$}
\end{quantikz}
\end{Code}
You can use standard \LaTeX\ maths expressions for your labels. Usually, spacing can be automatically adjusted just fine.

Sometimes, it might be that you want a multi-line label, and it should not be that each wire takes the height of those multiple lines. At this point, use the key \verb!disable auto height!. By default, each row will be assigned the height that a gate with label $U$ would be. This can be overridden by the third optional parameter of the gate command, if desired.

\begin{Code}[0.3]
\begin{quantikz}
\lstick{$c_0$} & \gate[3,disable auto height]{\verticaltext{MAJ}} & & \\
\lstick{$c_1$} &  &  \gate[3,disable auto height]{\verticaltext{MAJ}} & \\
\lstick{$c_2$} &  &  &  \\
\lstick{$c_3$} &  &  & 
\end{quantikz}
\end{Code}
Note that we used the command \verb!\verticaltext! for typesetting the text vertically.

Individual wires can also often be labelled. However, as this is typically an optional argument that appears late in the sequence, make sure you specify all proceeding parameters! For example,
\begin{Code}[0.3]
\begin{quantikz}
& \ctrl[wire style={"s"}]{1} & \wire[l][1]["p"{above,pos=0.2}]{a} \\
& \targ{} &
\end{quantikz}
\end{Code}
You will observe that stylings can also be passed. The \texttt{pos} is a fraction of the distance along the line. In this case, it's being drawn right to left, hence being closer to the right-hand edge\footnote{In the previous incarnation of quantikz, there was an accidental, undocumented method to label control wires. Internal changes have necessitated breaking that functionality.}.

\subsection{Boxing/Highlighting Parts of a Circuit}\label{sec:boxing}

It is often useful to highlight parts of a circuit. We do this with the \verb!\gategroup! command. The optional parameters \verb!wires! (the default) and \verb!steps! specify the number of rows and columns that the group spans respectively. The mandatory argument is the label for the box (although this can be empty). The top-left corner of the box coincides with the cell in which the command is placed.
\begin{Code}
\begin{quantikz}
& \gate{H} & \ctrl{1} & \gate{H}\gategroup[2,steps=3,style={inner sep=6pt}]{reversed c-{\sc not}} & \ctrl{1} & \gate{H} & \ctrl{1} & & \\
& & \targ{} & \gate{H} & \targ{} & \gate{H} & \targ{} & \gate{H} &
\end{quantikz}
\end{Code}
This is probably where you want to start to tune some of the optional parameters to style the box the way you want. For more details beyond some basic examples that follow, see Sec.\ \ref{sec:style}. By default, this box is drawn on top of the circuit itself. If you want it to be behind (for example, should you want it to have a background colour), then use the \verb!background! option.
\begin{Code}
\begin{quantikz}
& \gate{H} & \ctrl{1}\gategroup[2,steps=3,style={dashed,rounded corners,fill=blue!20, inner xsep=2pt},background,label style={label position=below,anchor=north,yshift=-0.2cm}]{{\sc swap}} & \targ{} & \ctrl{1} & & \\
& & \targ{} & \ctrl{-1} & \targ{} & \gate{H} &
\end{quantikz}
\end{Code}
The \verb!label style! key can be used to tune the label's properties, including positioning. Note that it is often good to use \verb!anchor=mid! for label anchors because if you have multiple labels, this will help get them horizontally aligned. It just means you have to use some \verb!yshift! to move the label off the border around the gategroup.

\subsection{Slicing}

It is often helpful to `slice' up a circuit for the sake of explaining it step by step. To do this, we provide the \verb!\slice{title}! command, which inserts a dashed vertical line after the column in which the command is added.
\begin{Code}
\begin{quantikz}
& \gate{H}\slice{step}  & \ctrl{1} & \meter{} \\
&       & \targ{} & \ctrl{1} & \gate{H} & \\
& & & \targ{} & &
\end{quantikz}
\end{Code}
You can also slice every step by using option \verb!slice all!, and the labels will be automatically numbered. This is likely to behave strangely unless you explicitly ensure that all rows have the same number of entries (i.e.\ short rows should have extra \& characters added).
\begin{Code}
\begin{quantikz}[slice all]
& \gate{H} & \ctrl{1} & \meter{} &\setwiretype{n}& \\
&       & \targ{} & \ctrl{1} & \gate{H} & \\
& & & \targ{} & &
\end{quantikz}
\end{Code}
If you need to adjust where the last slice is, use the optional parameter \verb!remove end slices!, which counts the number of columns fewer to add slices to. You can also change the title of each of the slices, by setting \verb!slice titles!. Include the macro \verb!\col! in your specification if you want to use the step number. Note, however, that the columns won't automatically space themselves out to accommodate a very wide label. You can style the slicing lines with the \verb!slice style! key, and the labels with \verb!slice label style!. These can be used to rotate the labels and create a bit more space!
\begin{Code}
\begin{quantikz}[slice all,remove end slices=1,slice titles=slice \col,slice style=blue,slice label style={inner sep=1pt,anchor=south west,rotate=40}]
& \gate{H} & \ctrl{1} & \meter{} &\setwiretype{n}& \\
&       & \targ{} & \ctrl{1} & \gate{H} & \\
& & & \targ{} & &
\end{quantikz}
\end{Code}
\noindent If you get compile errors when trying to slice, check the last line of your matrix, and make sure it doesn't end in \textbackslash\textbackslash.


\newpage
\section{Commands \& Options}
\reversemarginpar

\begin{description}[style=nextline]
\item [\textbackslash begin\{quantikz\}{[opts]}\ldots\textbackslash end\{quantikz\}]
The main environment in which you create quantum circuit diagrams. The main body of the environment is a table, with cells separated by \&, and new rows started by \textbackslash\textbackslash.

Note that each cell can contain at most one command indicated by the \faToggleOn\ symbol, and that command must be the first in the cell.

\texttt{opts} is a comma separated list of the following options:

\begin{tabular}{p{4cm}p{10cm}}
\texttt{wire types = \{list\}} & \texttt{list} is a comma separated list defining the wire type for each wire in the circuit, choosing from \texttt{q} (quantum), \texttt{c} (classical), \texttt{b} (wire bundle), \texttt{n} (none) If not specified, all wires are assumed to be quantum.\\
\texttt{thin lines} & Option for circuit aesthetic with thinner lines. \\
\texttt{transparent} & Sets entire circuit to have transparent background. \\
\texttt{classical gap=} & Sets \texttt{classical gap horizontal} and \texttt{classical gap vertical} to the specified value. \\
\texttt{classical gap vertical=} & Horizontal classical wires are created by placing two wires, at centre $\pm$ classical gap vertical. Default: 0.03cm. \\
\texttt{classical gap horizontal=} & Vertical classical wires are created by placing two wires, at centre $\pm$ classical gap horizontal. Default: 0.03cm. \\
\texttt{align equals at = n} & place the vertical centre of the circuit at wire number n (can be non-integer). If \verb!wire types! is specified, this is automatically set to $(N+1)/2$ where $N$ is the length of the list (i.e.\ number of wires).\\
\texttt{slice all} & add slices on all columns\\
\texttt{remove end slices = n} & does not show the last n slices\\
\texttt{slice titles =} & use the assigned text to label each slice. Use \verb!\col! to convey column number. \\
\texttt{slice style =}& Standard tikz style commands for the lines of the slice. Enclose in \{\} if giving multiple commands\\
\texttt{slice label style =} & Standard tikz style commands for the label of the slice. Enclose in \{\} if giving multiple commands\\
\texttt{vertical slice labels} & write the text of the slices vertically instead of horizontally. \\
\end{tabular}

The environment also receives any standard tikzcd (and tikz) options. See \href{http://mirrors.ctan.org/graphics/pgf/contrib/tikz-cd/tikz-cd-doc.pdf}{that manual} for more information. Particularly useful cases include:

\begin{tabular}{p{4cm}p{10cm}}
\texttt{column sep = n} & put spacing between each column of length n (e.g.\ 1cm). This is the padding between columns, \emph{not} the centres of gates, unless you add the \texttt{between origins} key.\\
\texttt{row sep = n} & put spacing between each row of length n (e.g.\ 1cm). This is the padding between rows, \emph{not} the centres of gates, unless you add the \texttt{between origins} key.\\
\texttt{between origins} & makes the row/column sep measurement be between the centres of gates. Usage: \texttt{row sep=\{0.6cm,between origins\}}
\end{tabular}

but you can also supply a colour, and that specifies the border colour of gates etc.

\item [\textbackslash setwiretype{[n]}\{t\}] 
Sets wire $n$ (default: current row, from that column onwards) to being of type $t$, which must be one of: \texttt{q} (quantum), \texttt{c} (classical) or \texttt{b} (bundle). Note that this will only affect the rendering from that point on, letting you change the wire type mid circuit. Should always come after any gate drawing command in a cell.
\begin{Code}
\begin{quantikz}
&&&\setwiretype{c}&&&
\end{quantikz}
\end{Code}
\item [\textbackslash wire{[d][n][s]}\{t\}]
Draw a wire starting from the current cell of type \texttt{t} which can be one of \texttt{a} (automatic), \texttt{q}, \texttt{c}, \texttt{b}, going in direction \texttt{d} (\texttt{u,d,l,r} for up/down/left/right) for a number of cells $n$. Automatic uses the current wire type of that wire. Note that the standard horizontal wire for that cell will still be drawn. Some styling can be specified using \texttt{s}. Largely unnecessary. Should always come after any gate drawing command in a cell.
\begin{Code}
\begin{quantikz}
&&&\wire[d][1]{c}&&&\\
&&&&&&
\end{quantikz}
\end{Code}
\item [\textbackslash wireoverride\{t\}]
Sets the wire type for the current cell (i.e.\ the wire starting in this cell and going backwards) to \texttt{t} without changing the wire type for the rest of the row. Should always come after any gate drawing command in a cell.
\begin{Code}
\begin{quantikz}
%of the 6 wire sections (count the &), knock out the third.
&&&\wireoverride{n}&&&
\end{quantikz}
\end{Code}
\item [\textbackslash qwbundle{[s]}\{n\}]
The standard typesetting of a wire bundle is 3 horizontal wires. Alternatively, you can use a single quantum wire, adding \textbackslash qwbundle to put a slash through the wire, labelled by \texttt{n}, typically the number of qubits that single wire represents. Should always come after any gate drawing command in a cell. The size of the strike can be altered by altering the parameter \texttt{s}:

\begin{tabular}{p{4cm}p{10cm}}
\texttt{style=} & Set the style of the line through a comma-separated list of tikz style commands. \\
\texttt{Strike Height=} & Set the height of the strike through\\
\texttt{Strike Width=} & Set the width of the strike through.
\end{tabular}

\begin{Code}
\begin{quantikz}[wire types={b,q},classical gap=0.07cm]
&&&& \\
& \qwbundle{n} &&&
\end{quantikz}
\end{Code}
\item [\textbackslash cbend\{a\}]\marginnote{\faToggleOn}[-10.5pt]
90 degree bend of a classical wire, running from start angle \texttt{a} (0: east, 90: north, 180: west, 270: south) to \texttt{a}+90. (Note that east to south is equivalent to south to east.)
\begin{Code}[0.35]
\begin{quantikz}[wire types={q,n,q}]
& \meter{}\wire[d][1]{c} &&& \\
& \cbend{0}&\setwiretype{c}&\cbend{180}&\setwiretype{n} \\
&&&\gate{U}\wire[u][1]{c}&
\end{quantikz}
\end{Code}
\item [\textbackslash permute\{list\}]
Permutation of wires. Takes the comma separated \texttt{list} of positive integers and connects wire $n$ to the wire specified in the $n^{th}$ element of the list. Counting starts such that wire 1 is the current wire. All wires are assumed to be quantum wires.

For example, \verb!\permute{3,1,2}! connects the first qubit to the third (relative to the location of the command, not absolute row number), the second to the first and the third to the second.
\begin{Code}
\begin{quantikz}[background color=black!5!white]
 &\permute{3,1,2} & \\
 &  &  \\
 &  & 
 \end{quantikz}
\end{Code}
Note that the wiring code is not sophisticated -- it will probably look quite ugly for anything complicated! Also, it assumes that the wires are quantum (not classical or bundled). Overlapping wires are given a gap by using a thicker line of the colour \texttt{background color} (default: white), so if your circuit sits on top of something non-white (as here), you'll need to change that.

\item [\textbackslash linethrough]\marginnote{\faToggleOn}[-10.5pt]
Objects (e.g.\ gates) in cells take up their own space that, by default, wires do not cross. This command puts a quantum wire across the current cell. Note that this will appear underneath multi-qubit gates, and therefore invisible unless those gates are transparent.

One place where this might be useful is as a ``pass-through'' on a gate, such as 
\begin{Code}
\begin{quantikz}[transparent]
& \gate[2]{J_{12}} & \gate[3,label style={yshift=0.2cm}]{J_{13}} & &   \\
& & \linethrough &\gate[2]{J_{23}} & \\
&&&&
\end{quantikz}
\end{Code}

\item [\textbackslash push\{t\}]
Places text \texttt{t} on the quantum wire, with no spacing or gate command around it.
\begin{Code}
\begin{quantikz}
&&\push{X}&&
\end{quantikz}
\end{Code}

\item [\textbackslash phantomgate\{s\}, \textbackslash hphantomgate\{s\}]\marginnote{\faToggleOn}[-10.5pt]
Places a quantum wire that occupies that same space as a single-qubit gate with label \texttt{s} would. The h version only occupies horizontal space, not vertical.
\begin{Code}
\begin{quantikz}
& \gate{H} & \phantomgate{really wide gate} & \gate{H} & 
\end{quantikz}
\end{Code}

\item [\textbackslash ghost{[w][h]}\{l\}] \marginnote{\faToggleOn}[-10.5pt]
Creates an invisible quantum gate that has the same height as \verb!\gate[][w][h]{l}!. Just like any other gate command, this should come before any other commands in a cell, and cannot be in the same cell as another gate command.
\begin{Code}
\begin{quantikz}
%give the wires the same vertical space as if they had H gates on them
&&\ghost{H} \\
&& \ghost{H}
\end{quantikz}=\begin{quantikz}
 & \gate{H} & \\
 & \gate{H} &
\end{quantikz}
\end{Code}

\item [\textbackslash phase{[s]}\{l\}]\marginnote{\faToggleOn}[-10.5pt]
Creates a phase gate (black circle) with label \texttt{l}. The optional parameter \texttt{s} controls the styling via the parameters

\begin{tabular}{p{4cm}p{10cm}}
\texttt{style=} & Set the style of the line through a comma-separated list of tikz style commands. \\
\texttt{label style=} & Set the style of the label through a comma-separated list of tikz style commands.
\end{tabular}

\begin{Code}
\begin{quantikz}
& \phase{\alpha} &
\end{quantikz}
\end{Code}

\item [\textbackslash lstick{[s]}\{t\}, \textbackslash midstick{[s]}\{t\}, \textbackslash rstick{[s]}\{t\}]
Text placement for labelling a wire to left/middle/right using text \texttt{t} (in text mode, not math mode) and styled using comma separated list of commands \texttt{s} from

\begin{tabular}{p{4cm}p{10cm}}
\texttt{n} & If $n$ is a positive integer value, this is interpreted as the number of wires to use (default is 1). By default, introduces curly braces to span multiple wires. \\
\texttt{wires=n} & long-hand form of the above.\\
\texttt{label style=} & Styles the text using standard tikz commands. If using multiple commands, or anything involving a space, enclose in \{\}.\\
\texttt{brackets=}& Choose from \texttt{none/left/right/both} to switch on/off appropriate bracketing. (left won't affect \textbackslash lstick, and right won't affect \textbackslash rstick). \\
\texttt{braces=} & Styles the braces using standard tikz commands.
\end{tabular}

\begin{Code}
\begin{quantikz}
\lstick{\ket{0}} & \gate{H} & \midstick[2]{middle} & \gate{H}& \rstick[2,brackets=none]{out} \\
\lstick{\ket{0}} & \gate{H} &&\gate{H}&
\end{quantikz}
\end{Code}

\item [\textbackslash gate{[opts][w][h]}\{l\}] \marginnote{\faToggleOn}[-10.5pt]
The standard quantum gate command. Creates a box containing the label \texttt{l}. \texttt{w} and \texttt{h} are optional minimum width and minimum height parameters to override the automatic sizing. Gate is styled using the optional parameter \texttt{opts}, which is a comma separated list of commands:

\begin{tabular}{p{4cm}p{10cm}}
\texttt{n} & If $n$ is a positive integer value, this is interpreted as the number of wires to use (default is 1).\\
\texttt{wires=n} & long-hand form of the above.\\
\texttt{style=} & Styles the gate (box) using standard tikz commands. If using multiple commands, or anything involving a space, enclose in \{\}.\\
\texttt{label style=} & Styles the gate text using standard tikz commands. If using multiple commands, or anything involving a space, enclose in \{\}.\\
\texttt{disable auto height} & pretty self-explanatory\\
\texttt{swap} & Fixes the gate to be a 2-qubit gate, depicting a swap between the two wires.
\end{tabular}

\begin{Code}
\begin{quantikz}
& \gate{H} & \gate[2]{U} & \gate[2,swap]{} & \\
&&&&
\end{quantikz}
\end{Code}

\item [\textbackslash gateinput{[s]}\{l\}, \textbackslash gateoutput{[s]}\{l\}]
Put a label \texttt{l} inside the current gate command, starting on the current row, on either the input or output. If spanning multiple rows, will group the wires using curly braces by default. The width of the containing gate does not automatically adjust to the contents of these extra labels, so you will have to add it with the second optional parameter of \verb!\gate!. Style with parameter \texttt{s} using comma separated list of commands from:

\begin{tabular}{p{4cm}p{10cm}}
\texttt{n} & If $n$ is a positive integer value, this is interpreted as the number of wires to span (default is 1).\\
\texttt{wires=n} & long-hand form of the above.\\
\texttt{label style=} & Styles the text using standard tikz commands. If using multiple commands, or anything involving a space, enclose in \{\}.\\
\texttt{braces=} & Styles the braces using standard tikz commands.
\end{tabular}

\begin{Code}[0.3]
\begin{quantikz}
&\ctrl{1} & \\
&\gate[3][1.7cm]{U}\gateinput[2]{$x$}\gateoutput[2]{$x$} & \\
& & \\
&\gateinput{$y$}\gateoutput{$y\oplus f(x)$}
&
\end{quantikz}
\end{Code}

\item [\textbackslash meter{[opts][w][h]}\{l\}, \textbackslash metercw{[opts][w][h]}\{l\}]\marginnote{\faToggleOn}[-10.5pt]
Measurement gates of different styles. Measurement is labelled by label \texttt{l}. Styling specified by optional parameter \texttt{opts}, a comma separated list of tikz commands. (Note the similarity to the gate command.) Can span multiple wires.

\begin{Code}[0.3]
\begin{quantikz}
& \metercw[label style={inner sep=1pt}]{x} & \meter[2]{y} \\
&&
\end{quantikz}
\end{Code}

\begin{tabular}{p{4cm}p{10cm}}
\texttt{n} & If $n$ is a positive integer value, this is interpreted as the number of wires to use (default is 1).\\
\texttt{wires=n} & long-hand form of the above.\\
\texttt{style=} & Styles the gate (box) using standard tikz commands. If using multiple commands, or anything involving a space, enclose in \{\}.\\
\texttt{label style=} & Styles the gate text using standard tikz commands. If using multiple commands, or anything involving a space, enclose in \{\}.\\
\texttt{disable auto height} & pretty self-explanatory\\
\end{tabular}

\item [\textbackslash measure{[s]}\{l\}, \textbackslash measuretab{[s]}\{l\}, \textbackslash meterD{[s]}\{l\}, \textbackslash inputD{[s]}\{l\}]\marginnote{\faToggleOn}[-10.5pt]
Single-qubit measurement gates of different styles. Measurement is labelled by label \texttt{l}. Styling specified by optional parameter \texttt{s}, a comma separated list of tikz style commands. Note that this is provided as \verb![s]! where most other gates would expect \verb![style=s]!. \verb!\inputD! is the equivalent for inputs of \verb!\meterD!. For \verb!\inputD!, incoming wire is disabled, overriding global style.
\begin{Code}
\begin{quantikz}
& \measure{x} \\
& \measuretab{x} \\
\inputD{0} & \meterD{x\vphantom{0}}
%vphantom makes the input and output shapes the same height, because x and 0 are different heights
\end{quantikz}
\end{Code}

\item [\textbackslash meterout\{l\}]
Typically used immediately after a \verb!\meter! command, gives a right-pointing arrow entering the cell, up to text \texttt{l} (in math mode). The wire type for the rest of the row is set to `none'.
\begin{Code}
\begin{quantikz}
& \meter{}\wire[d][2]{c} & \meterout{a} \\
& \meter{} & \meterout{b} \\
& \gate{G^{a\oplus b}} &
\end{quantikz}
\end{Code}

\item [\textbackslash swap{[s]}\{n\}] \marginnote{\faToggleOn}[-10.5pt]
Create an X shape on the current wire and a single vertical wire going downwards $n$ rows (may be negative). Used as a swap gate, typically paired with \verb!\targX!. Styled using parameter \texttt{s} as comma separated list of commands, some of which provide access to a partial swap:

\begin{tabular}{p{4cm}p{10cm}}
\texttt{partial swap=}&The text to place inside a circle \\
\texttt{partial position=}&Fractional position of the circle along the vertical line. Default: 0.5 (i.e.\ middle). \\
\texttt{style=}&Tikz styling parameters for the gate. \\
\texttt{label style=}&Tikz styling parameters for the label, overriding those of the whole gate. \\
\texttt{vertical wire=}& Specifies the type of vertical wire: \texttt{q} (quantum, default), \texttt{c} (classical) or \texttt{b} (bundle)
\end{tabular}

\begin{Code}
\begin{quantikz}[row sep=0.8cm]
& \swap[partial swap={\pi/4},partial position=0.3]{2} & \swap{1} &  \\
& & \targX{} & \\
& \targX{} &&
 \end{quantikz}
\end{Code}

\item [\textbackslash ctrl{[s]}\{n\}, \textbackslash octrl{[s]}\{n\}]\marginnote{\faToggleOn}[-10.5pt]
Starting point of a controlled gate, using filled circle (\textbackslash ctrl) or open circle (\textbackslash octrl). By default, single vertical line is added to go down $n$ rows (may be negative). Settings are controlled via the optional parameter \texttt{s}. \textbackslash ctrl\{\} is a synonym for \textbackslash control\{\}. Thus everything you need can be accomplished just with the \textbackslash ctrl command, without resorting to all the variants.

\begin{tabular}{p{4cm}p{10cm}}
\texttt{style=}&Tikz styling parameters for the gate (both the circle and the vertical wire).\\
\texttt{wire style=}& Styling parameters for the wire only, overriding anything set by \texttt{style}. \\
\texttt{vertical wire=}& Specifies the type of vertical wire: \texttt{q} (quantum, default), \texttt{c} (classical) or \texttt{b} (bundle) \\
\texttt{open}& Makes the circle an open circle, i.e.\ \verb!\ctrl[open]{1}! is a synonym for \verb!\octrl{1}!.
\end{tabular}

\begin{Code}
\begin{quantikz}
& \ctrl{2} & \octrl{1} & \\
& \control{} & \targ{} & \\
& \ocontrol{} &&
\end{quantikz}
\end{Code}

\item [\textbackslash control{[s]}\{\}, \textbackslash ocontrol{[s]}\{\}]\marginnote{\faToggleOn}[-10.5pt]
Target equivalents of the above commands, but with no vertical wire. See previous two entries for usage examples.

\begin{tabular}{p{4cm}p{10cm}}
\texttt{style=}&Tikz styling parameters for the gate (both the circle and the vertical wire).\\
\texttt{open}& Makes the circle an open circle, i.e.\ \verb!\control[open]{}! is a synonym for \verb!\ocontrol{}!.
\end{tabular}

\item [\textbackslash targ{[s]}\{\}, \textbackslash targX{[s]}\{\}]\marginnote{\faToggleOn}[-10.5pt]
Target elements of controlled-not and controlled-swap (no vertical wire). \texttt{s} provides styling parameters. See previous entries for usage examples.

\begin{tabular}{p{4cm}p{10cm}}
\texttt{style=}&Tikz styling parameters for the gate.
\end{tabular}

\item [\textbackslash gategroup{[opts]}\{l\}]
Create a large box with top-left-hand corner positioned (roughly) at the top-left of the current cell. The box has label \texttt{l} and is styled with the options \texttt{opts}:

\begin{tabular}{p{4cm}p{10cm}}
\texttt{n} & If $n$ is a positive integer value, this is interpreted as the number of rows (wires) to cover (default is 1).\\
\texttt{wires=n} & long-hand form of the above.\\
\texttt{steps=m} & This is the number of columns (time steps) that the box covers. Default is 1. \\
\texttt{style=} & Styles the box using standard tikz commands. If using multiple commands, or anything involving a space, enclose in \{\}.\\
\texttt{label style=} & Styles the label text using standard tikz commands. If using multiple commands, or anything involving a space, enclose in \{\}. \\
\texttt{background} & Draw the gategroup behind the circuit (useful if the box has a background colour).
\end{tabular}

\begin{Code}
\begin{quantikz}
 & \ctrl{1} & \gate{H}\gategroup[2,steps=3,style={inner sep=6pt}]{reversed c-{\sc not}} & \ctrl{1} & \gate{H} & \ctrl{1} & \\
 & \targ{} & \gate{H} & \targ{} & \gate{H} & \targ{} &
\end{quantikz}
\end{Code}

\item [\textbackslash slice{[s]}\{l\}]
Insert a slice between the current column and the next, with a title of \texttt{l}. Styles can be applied via the optional parameter \texttt{s}:

\begin{tabular}{p{4cm}p{10cm}}
\texttt{style=}&Tikz styling parameters for the slice. \\
\texttt{label style=}&Tikz styling parameters for the title/label.
\end{tabular}

\begin{Code}
\begin{quantikz}
& \gate{H}\slice{step}  & \ctrl{1} & \meter{} \\
&       & \targ{} & \ctrl{1} & \gate{H} & \\
& & & \targ{} & &
\end{quantikz}
\end{Code}

\item [\textbackslash makeebit{[s]}\{l\}]
Create an e-bit. This places the label \texttt{l} halfway between the current row and the next row. Two lines come off this label, leading to the current wire and the one below.

\begin{tabular}{p{4cm}p{10cm}}
\texttt{style=}&Tikz styling parameters for the wires. \\
\texttt{label style=}&Tikz styling parameters for the label. \\
\texttt{angle=}&The angle of the two lines drawn, in degrees. Default: -45
\end{tabular}

\begin{Code}
\begin{quantikz}
\makeebit[angle=-60,label style=blue]{generate} & & \\
 & &
\end{quantikz}
\end{Code}

\item [\textbackslash trash{[s]}\{l\}, \textbackslash ground{[s]}\{l\}] \marginnote{\faToggleOn}[-10.5pt]
Two different ways of denoting the termination of a wire (e.g.\ tracing out). Label text \texttt{l}, styling options directly supplied via \texttt{s}.
\begin{Code}
\begin{quantikz}
&\gate[3]{U} & \\
&& \ground{} \\
&& \trash{\text{trace}}
\end{quantikz}
\end{Code}

\item [\textbackslash wave{[s]}\{\}]
Draw a wave along an entire row. \texttt{s} is standard tikz formatting commands for additional control over the style. By default, there is no wire drawn on this row. To have one, you should run \textbackslash setwiretype\{q\} immediately after the \textbackslash wave command.

\begin{Code}
\begin{quantikz}
& \gate{H} & \ctrl{3} & \ \ldots\  & & \gate{H} & \\
\wave{}&&&&&&\\
& \gate{H} &  & \ \ldots\  & \ctrl{1} & \gate{H} & \\
& & \gate{U} & \ \ldots\  & \gate{U^k} &  & 
\end{quantikz}
\end{Code}

\item [\textbackslash verticaltext\{l\}]
Present the text in label \texttt{l} as vertically stacked. Can be helpful for slices.

\item [\textbackslash ket\{l\}, \textbackslash bra\{l\}, \textbackslash proj\{l\}, \textbackslash braket\{l\}\{m\}]
Typeset Dirac notation $\ket{l}$, $\bra{l}$, $\proj{l}$ and $\braket{l}{m}$ respectively. These commands do not require math mode, and the braces will automatically resize to the argument. They are defined to behave well with other packages (e.g.\ physics) that may define the same commands. You may need to be careful of the order in which you load those packages: load quantikz \emph{after} the other package --- if quantikz sees that those commands are already defined, it does not redefine them. If you wish to ensure that you are using the version that this package defines, run \textbackslash forceredefine at the end of your preamble (after all packages have loaded). 

\end{description}

\section{Spacing}\label{sec:spacing}

\subsection{Local Adjustment}

There are several different ways in which we can manipulate the spacing of a diagram. Adding space to an individual row or column can be done in the standard way of tables in LaTeX. Here we add 2cm of space to the column between the $H$ and $X$ gates, and 1cm of space between the top two rows.
\begin{Code}
\begin{quantikz}
& \gate{H} &[2cm] \gate{X} & \gate{H} & \\[1cm]
& \gate{X} & \gate{Z} & \gate{Z} &\\
& \gate{X} & \gate{Z} & \gate{Z} &
\end{quantikz}
\end{Code}
If you don't know how much space you need, but it should be determined by the size of some text, you can use \verb!\hphantom{}! (widens the gate, in a similar way to \verb!\gate[1cm]{}!) or \verb!\hphantomgate{}! (increases the length of a wire) for horizontal spacing, and \verb!\ghost{}! for vertical spacing.
\begin{Code}
\begin{quantikz}
& \gate{X} \hphantom{very wide} & \gate{X} & \hphantomgate{wide} & \gate{X} &
\end{quantikz}
\end{Code} 

\subsection{Global Adjustment}

Standard tikz commands facilitate a global adjustment of row and column spacing. For example, a ridiculous horizontal spacing:
\begin{Code}
\begin{quantikz}[column sep=1cm]
& \gate{H} & \phase{\beta} & \gate{H} &
\end{quantikz}
\end{Code}
This specifically adjusts the \emph{gap} between the rows and columns, not the distance between the centres of the rows and columns. Depending on what gates you have on each wire, the spacing may not be the same between each wire. Sometimes this is desirable, particularly if a gate in one particular row is much larger than anything in the other rows. At other times, it just makes your diagram look a little odd. For example, look at the gap between the top two wires and the bottom two wires:
\begin{Code}
\begin{quantikz}[row sep=0.1cm]
& \gate{X} & \ctrl{1} & \gate{X} & \\
& & \ctrl{} &  & \\
& \gate{X} & & & \\
& \gate{H} & & & 
\end{quantikz}
\end{Code}
If you want to make sure that every quantum wire is equally spaced, do the following to \verb!row sep!:
\begin{Code}
\begin{quantikz}[row sep={0.6cm,between origins}]
& \gate{X} & \ctrl{1} & \gate{X} & \\
& & \ctrl{} & & \\
& \gate{X} & & & \\
& \gate{H} & & & 
\end{quantikz}
\end{Code}
This is particularly useful to achieve alignment of several circuits, as in \ref{sec:align}.

\subsection{Alignment}\label{sec:align}

How do we centre a circuit diagram? Simply surround it with \verb!\begin{center}! and \verb!\end{center}! commands, or within any standard equation environment.

Vertical alignment between different circuits can be more fiddly, depending on how much of a perfectionist you are. Sometimes, they work immediately, but the wires don't always align perfectly with each other. Generally the problem is that the highest gate in each row is different (here, the LHS is missing an $X$ gate on the second row)
\begin{Code}
\begin{quantikz}
& \gate{X} & \ctrl{1} & \\
& & \targ{} &
\end{quantikz}
=\begin{quantikz}
& \ctrl{1} & \gate{X} & \\
& \targ{}  & \gate{X} &
\end{quantikz}
\end{Code}
\noindent  Ensuring an even spacing between rows, as described in Sec.\ \ref{sec:spacing}, can help (but is not always appropriate). Often the easiest is to fudge it using the \verb!\ghost! command which will add a 0-width gate of the height corresponding to its argument. So, having identified the problem with the above circuit, we can replace it with:
\begin{Code}
\begin{quantikz}
& \gate{X} & \ctrl{1} & \\
& \ghost{X} & \targ{} &
\end{quantikz}
=\begin{quantikz}
& \ctrl{1} & \gate{X} & \\
& \targ{}  & \gate{X} &
\end{quantikz}
\end{Code}

If you cannot identify the offending gate, and particularly if the operation is not a standard \verb!\gate! command, you might be better off combining the two circuits in a single circuit with no wires joining the two parts. You can use a \texttt{midstick} command here. By default, it places braces both before and after, but these can be replaced using the optional argument \texttt{brackets=none|left|right|both}. Thus,
\begin{Code}
\begin{quantikz}
& & \ctrl{1} & \midstick[2,brackets=none]{=}& \ctrl{1} & \gate{Z} & \\
& \gate{Z} & \targ{} & & \targ{}  & \gate{Z} &
\end{quantikz}
\end{Code}

 \subsubsection{Perfecting Vertical Alignment}

Sometimes when you're typesetting circuit identities as multiple separate circuits, the vertical alignment of the equals sign doesn't appear quite right (and can really niggle). Here, for example, the equals seems a bit low (because the baseline is the middle of the diagram by default, and here, the rows are not equal heights):
\begin{Code}
\begin{quantikz}
&& \\
& \gate{Z} &
 \end{quantikz}=\begin{quantikz}
&&&& \\
& \gate{H} & \gate{X} & \gate{H} &
 \end{quantikz}
\end{Code}
To that end, we have added the key \texttt{align equals at=} option for the quantikz environment. This specifies which wire should be aligned with the equals sign. You can even use a non-integer. For instance, 1.5 will set it half way between wires 1 and 2.
\begin{Code}
\begin{quantikz}[align equals at=1.5]
&& \\
& \gate{Z} &
 \end{quantikz}=\begin{quantikz}[align equals at=1.5]
&&&& \\
& \gate{H} & \gate{X} & \gate{H} &
 \end{quantikz}
\end{Code}
If you use the \texttt{wire types} global option, this happens automatically. You can still override it provided the \texttt{align equals at} comes \emph{after}.

\section{Styling \& Customising}\label{sec:style}

\subsection{Global Styling}

If you want to change the properties of an entire circuit such that all the typically black elements are a different colour, and the backgrounds of cells are another, you can supply the quantikz command with two keys: \verb!color! and \verb!background color!. The second of these works well alongside the \verb!\pagecolor! command for making the rest of the page a particular colour.
\begin{Code}
\begin{quantikz}[color=blue,background color=yellow]
\lstick{\ket{\psi}} & \gate{H} & \gate{X} & \meter{}
\end{quantikz}
\end{Code}

There are two further keys that change styles globally: `thin lines' to make the lines thin, more in keeping with what QCircuit produced, and `transparent', should you want the background of all the gates to be transparent:
\begin{Code}[0.45]
\begin{quantikz}[thin lines,transparent]
& \ctrl{1} & & \ctrl{1} & & \\
& \targ{} & \gate{R_z(-\theta/2)} & \targ{} & \gate{R_z(\theta/2)} & \meter{}
\end{quantikz}
\end{Code}

Global properties that affect all circuit elements of a given type can be affected through \verb!tikzset!.
\begin{Code}
\tikzset{
 operator/.append style={fill=red!20},
 my label/.append style={above right,xshift=0.3cm},
 phase label/.append style={label position=above}
}
\begin{quantikz}
& \gate{H} & \phase{\beta} & \gate{H} & \meter{\ket{\pm}}
\end{quantikz}
\end{Code}
The global styles are:
\begin{center}
\begin{tabular}{c|c}
Style Name & Affected Command(s)  \\
\hline
operator & \verb!\gate!  \\
meter & \verb!\meter! \\
slice & \verb!\slice! \\
wave & \verb!\wave! \\
leftinternal & \verb!\gateinput! \\
rightinternal   & \verb!\gateoutput! \\
dm & left braces (\verb!\gateoutput!,\verb!\lstick!) \\
dd & right braces (\verb!\gateinput!,\verb!\rstick!) \\
phase & \verb!\phase!, \verb!\control!, \verb!\ophase!, \verb!\ocontrol! \\
circlewc & \verb!\targ! \\
crossx2 & \verb!\swap!,\verb!\targX! \\
my label & measurement bases in \verb!\meter! \\
phase label & phases in \verb!\phase! \\
gg label & main gate label in \verb!\gate! \\
group label & label in \verb!\gategroup!
\end{tabular}
\end{center}

If the inconsistency of gate heights annoys you for instance,
\begin{Code}
\begin{quantikz}
& \gate{e} & \gate{L^1_2} &
\end{quantikz}
\end{Code}
then fix it by giving every gate a suitable \texttt{minimum height}:
\begin{Code}
\tikzset{
 operator/.append style={minimum height=0.7cm}
 }
\begin{quantikz}
& \gate{e} & \gate{L^1_2} &
\end{quantikz}
\end{Code}

\subsection{Per-Gate Styling}

Individual gates can be modified using optional arguments of the calling function. 
\begin{Code}
\begin{quantikz}
& \gate[style={fill=red!20},label style=cyan]{H} & \phase[style={green},label style={label position=above}]{\beta} & \gate{H} & & \meter[style={draw=blue}]{\ket{\pm}}
\end{quantikz}
\end{Code}
The specific syntax varies a little depending on the type of gate. See the commands list for the specific cases.
If you want to input several styling parameters with one of the keys, just group them together in a set of curly braces, \{\}. Typical styling parameters include \verb!draw=! specifying line colour, \verb!fill=!, specifying fill colour, \verb!inner xsep=! and \verb!inner ysep=! specifying horizontal and vertical margins respectively, \verb!xshift=! and \verb!yshift=! for adjusting horizontal and vertical positioning. Beyond that, the \href{http://mirrors.ctan.org/graphics/pgf/base/doc/pgfmanual.pdf}{tikz manual} is your friend!

This styling is really quite flexible, as we can override the default shapes with anything that we want:
\begin{Code}
\tikzset{
 noisy/.style={starburst,fill=yellow,draw=red,line width=2pt,inner xsep=-4pt,inner ysep=-5pt}
}
\begin{quantikz}[row sep=0.3cm,column sep=0.3cm,wire types={q,q,q,n,n}]%
& \gate{H} & \ctrl{2}& & \gate[3,style={noisy},label style=cyan]{\text{noise}} & \ctrl{3} &  & & \\
\lstick{\ket{0}} & & \targ{} & & & & \ctrl{3} & & \\
\lstick{\ket{0}} & & \targ{}& & & & & \ctrl{2} & \\
&&&&\lstick{\ket{0}} & \targ{}\setwiretype{q} & \targ{} & & \meter{} \\
&&&&\lstick{\ket{0}} & \setwiretype{q} & \targ{} & \targ{} & \meter{} 
\end{quantikz}
\end{Code}

\subsection{Scaling}

If you want to override the standard size of a circuit (gate elements, text and spacing), you can make it a node inside a \verb!tikzpicture!:
\begin{Code}
\begin{tikzpicture}
\node[scale=1.5] {
\begin{quantikz}
& \gate{H} & \phase{\beta} & \gate{H} &
\end{quantikz}
};
\end{tikzpicture}
\end{Code}
It's also possible to rescale to a fixed width, so long as you declare the \verb!adjustbox! package in the document preamble.
\begin{Code}
\begin{adjustbox}{width=0.8\textwidth}
\begin{quantikz}
& \gate{H} & \phase{\beta} & \gate{H} &
\end{quantikz}
\end{adjustbox}
\end{Code}

\section{Bells and Whistles}

Since we have built quantikz on top of tikzcd, any of the standard arrow commands will work (don't forget to turn off the default wire!). For example, after a measurement, you might want to use an arrow to report a particular measurement outcome using \verb!\arrow[r]!. The \verb!r! conveys that the arrow should head one cell to the right. You can use combinations of up (u), down (d), left (l) and right as you wish. For more styling options, see the \href{http://mirrors.ctan.org/graphics/pgf/contrib/tikz-cd/tikz-cd-doc.pdf}{tikzcd manual}.
\begin{Code}
\begin{quantikz}
\lstick{\ket{0}\\initial state} & & \push{X} & & \meter{0/1} \arrow[r] & \rstick{\ket{1}}\setwiretype{n}
\end{quantikz}
\end{Code}

It's perhaps worth mentioning that gate commands can include matrices.
\begin{Code}
\begin{quantikz}
\lstick{\ket{0}} & \gate{\frac{1}{\sqrt{2}}\begin{bmatrix} 1 & 1 \\ 1 & -1 \end{bmatrix}} & \rstick{\ket{+}}
\end{quantikz}
\end{Code}

Tikzcd provides a very useful global key called \texttt{execute at end picture} which allows you to execute a set of commands after the drawing of the circuit has completed (just make sure that you don't include any blank lines in the code!). This means that you can refer, for example, to the positions of any cell. If you want the cell at row n and column n, just use \texttt{\textbackslash tikzcdmatrixname-n-m}. By default, labels sit outside the standard spacing of the underlying matrix, and do not have names. You can give a label a name (such as `Alabel' in the example below) so that you can also refer to it. Also remember that different cells, with different elements, will have different borders. That's why it can help to define a new node (e.g.\ `toprow') that makes sure all the elements you want \texttt{fit} inside it. In fact, there are already blocks defined that group all elements in rows and columns, which you can access as \texttt{\textbackslash tikzcdmatrixname-rown} and \texttt{\textbackslash tikzcdmatrixname-colm}
\vfill

\begin{Code}[0.3]
\begin{quantikz}[align equals at=2,execute at end picture = {%
    \node (toprow) [fit=(\tikzcdmatrixname-1-3) (\tikzcdmatrixname-1-2)] {};
    \node (firstcol) [fit=(\tikzcdmatrixname-1-2) (\tikzcdmatrixname-2-2),inner sep=6pt] {};
    \node (nextrow) [fit=(\tikzcdmatrixname-2-1) (\tikzcdmatrixname-2-2) (Alabel),inner sep=0pt] {};
    \draw [thick,draw=none,dashed,rounded corners,fill=red,fill opacity=0.2] 
    (toprow.north east) -- (toprow.south east) -- (firstcol.east|-toprow.south) -- (firstcol.east|-nextrow.south) -- (nextrow.south west) -- (nextrow.north west) -- (firstcol.west|-nextrow.north) -- (firstcol.west|-toprow.north) -- cycle;
%
%
    \draw [thick,draw=none,dashed,rounded corners,fill=red,fill opacity=0.2] 
    (\tikzcdmatrixname-col4.east|-\tikzcdmatrixname-row2.north) -- (\tikzcdmatrixname-col4.east|-\tikzcdmatrixname-row2.south) -- 
    (\tikzcdmatrixname-col3.east|-\tikzcdmatrixname-row2.south) -- (\tikzcdmatrixname-col3.east|-Blabel.south) -- 
    (Blabel.south west) -- (Blabel.north west) -- (\tikzcdmatrixname-col3.west|-Blabel.north) -- 
    (\tikzcdmatrixname-col3.west|-\tikzcdmatrixname-row2.north) -- cycle;
}% end of execute at end picture
]
\lstick{\ket{\psi_{\text{in}}}} & \ctrl{1} & \meterD{\ket{\pm}}    \\
\lstick[label style={name=Alabel}]{\ket{+}} & \ctrl{}  &  \ctrl{1} & \meterD{\ket{\pm}}  \\
\lstick[label style={name=Blabel}]{\ket{+}} &  & \ctrl{}  & &[-0.5cm]
\end{quantikz}
\end{Code}

\subsection{Externalization}

A large document with many quantum circuits in it can be very slow to compile. This is where externalization comes in, producing images for each circuit. (It's also helpful if you have to produce accessible versions of lecture notes!) Quantikz \emph{should} work with the externalization routines of tikz. Turn it on just as you normally would,
\begin{FullCode}
\begin{lstlisting}
\usetikzlibrary{external}
\tikzexternalize % activate!
\end{lstlisting}
\end{FullCode}
Note that for any circuits within an amsmath environment such as align, since these are run twice, you may end up with two copies of a given image.

\subsection{New Gate Types}

One of the most common types of email I get about quantikz goes along the lines of ``Thank you for this brilliant package. I really want to use it, but I need \emph{this} specific gate shape that nobody else has ever heard of before for my own personal reasons and I can't believe you haven't implemented it already. Fix it now!''. This is impractical. However, with the latest version of quantikz, you can now use standard tikz commands in order to define your own shapes for single-qubit gates. Be warned: it can be quite fiddly.

The key to the process is defining your own shape (or appropriating an existing shape from the \href{https://tikz.dev/library-shapes}{shapes library}). As an example, we will use the \texttt{isosceles triangle} shape.

Then we need to ensure that wires will join on to your shape at the correct places. There are two different mechanisms for this. Quantum wires use the standard shape border, i.e., if you draw a straight line between the centre of two cells, the point of the join should be where this line intersects the edge of your shape. For standard shapes, this will typically just work If you're specifying your own shape, you'll need to ensure that the function \verb!\anchorborder! function works correctly, at least at the north/south/east/west points. Classical wires (and bundles) attach to different anchors on the edge of the shape. We need 16 of them (but really, 8 are just aliases of the first 8). For example, a left-going classical wire will involve drawing two horizontal lines. One starts at \texttt{lstartone} and ends at the \texttt{lendone} of the target node. The other starts at \texttt{lstarttwo} and ends at \texttt{lendtwo}. The \texttt{lstartone} anchor should be on the left-most edge of your shape, at a height \pgfkeysvalueof{/tikz/commutative diagrams/classical gap vertical} above the center, but then you will have to calculate the correct horizontal position. Similarly, \texttt{lstarttwo} should be on the left-most edge of your shape, at a height \pgfkeysvalueof{/tikz/commutative diagrams/classical gap vertical} below the center. The end points are the equivalent points on the right-hand edge, and will also correspond to \texttt{rstartone} and \texttt{rstarttwo}.
\begin{FullCode}
\begin{lstlisting}
\usetikzlibrary{shapes.geometric}
\makeatletter

%add new anchors
\pgfaddtoshape{isosceles triangle}{
%define the anchor lstartone.
  \anchor{lstartone}{%
    \trianglepoints %built-in function for this shape. Defines certain macros with position info
    %get the position of the lower left corner. Stored in macros \pgf@x and \pgf@y
    \pgf@process{\pgfpointadd{\centerpoint}{\pgfmathrotatepointaround{\lowerleftanchor}{\pgfpointorigin}{\rotate}}}
    \pgf@ya=.5\pgf@y
    \pgf@process{\pgfpointadd{\centerpoint}{\pgfmathrotatepointaround{\lowerrightanchor}{\pgfpointorigin}{\rotate}}}
    \pgf@y=.5\pgf@y
    \advance\pgf@y by \pgf@ya%
    \advance\pgf@y by -\pgfkeysvalueof{/tikz/commutative diagrams/classical gap vertical}%
    %by end of function, position of anchor stored in \pgf@x and \pgf@y
  }
  \anchor{lstarttwo}{%
    \trianglepoints
    \pgf@process{\pgfpointadd{\centerpoint}{\pgfmathrotatepointaround{\lowerleftanchor}{\pgfpointorigin}{\rotate}}}
    \pgf@ya=.5\pgf@y
    \pgf@process{\pgfpointadd{\centerpoint}{\pgfmathrotatepointaround{\lowerrightanchor}{\pgfpointorigin}{\rotate}}}
    \pgf@y=.5\pgf@y
    \advance\pgf@y by \pgf@ya%
    \advance\pgf@y by \pgfkeysvalueof{/tikz/commutative diagrams/classical gap vertical}%
  }
  \anchor{lendone}{%
    \trianglepoints
    \pgf@process{\pgfpointintersectionoflines%
      {\pgfpointadd{\centerpoint}{\pgfmathrotatepointaround{\apexanchor}{\pgfpointorigin}{\rotate}}}%
      {\pgfpointadd{\centerpoint}{\pgfmathrotatepointaround{\lowerrightanchor}{\pgfpointorigin}{\rotate}}}%
      {\pgfpointadd{\centerpoint}{\pgfpoint{0cm}{-\pgfkeysvalueof{/tikz/commutative diagrams/classical gap vertical}}}}%
      {\pgfpointadd{\centerpoint}{\pgfpoint{0.5cm}{-\pgfkeysvalueof{/tikz/commutative diagrams/classical gap vertical}}}}%
    }%
  }
  \anchor{lendtwo}{%
    \trianglepoints
    \pgf@process{\pgfpointintersectionoflines%
      {\pgfpointadd{\centerpoint}{\pgfmathrotatepointaround{\apexanchor}{\pgfpointorigin}{\rotate}}}%
      {\pgfpointadd{\centerpoint}{\pgfmathrotatepointaround{\lowerleftanchor}{\pgfpointorigin}{\rotate}}}%
      {\pgfpointadd{\centerpoint}{\pgfpoint{0cm}{\pgfkeysvalueof{/tikz/commutative diagrams/classical gap vertical}}}}%
      {\pgfpointadd{\centerpoint}{\pgfpoint{0.5cm}{\pgfkeysvalueof{/tikz/commutative diagrams/classical gap vertical}}}}%
    }%
  }
  \end{lstlisting}
  \end{FullCode}
  Similarly, we need 4 anchors for the down wire.
  \begin{FullCode}
  \begin{lstlisting}
  %now do the anchors for a down wire
  \anchor{dstarttwo}{%
    \trianglepoints
    \pgf@process{\pgfpointintersectionoflines%
      {\pgfpointadd{\centerpoint}{\pgfmathrotatepointaround{\apexanchor}{\pgfpointorigin}{\rotate}}}%
      {\pgfpointadd{\centerpoint}{\pgfmathrotatepointaround{\lowerrightanchor}{\pgfpointorigin}{\rotate}}}%
      {\pgfpointadd{\centerpoint}{\pgfpoint{\pgfkeysvalueof{/tikz/commutative diagrams/classical gap horizontal}}{0cm}}}%
      {\pgfpointadd{\centerpoint}{\pgfpoint{\pgfkeysvalueof{/tikz/commutative diagrams/classical gap horizontal}}{0.5cm}}}%
    }%
  }
  \anchor{dstartone}{%
    \trianglepoints
    \pgf@process{\pgfpointintersectionoflines%
      {\pgfpointadd{\centerpoint}{\pgfmathrotatepointaround{\apexanchor}{\pgfpointorigin}{\rotate}}}%
      {\pgfpointadd{\centerpoint}{\pgfmathrotatepointaround{\lowerrightanchor}{\pgfpointorigin}{\rotate}}}%
      {\pgfpointadd{\centerpoint}{\pgfpoint{-\pgfkeysvalueof{/tikz/commutative diagrams/classical gap horizontal}}{0cm}}}%
      {\pgfpointadd{\centerpoint}{\pgfpoint{-\pgfkeysvalueof{/tikz/commutative diagrams/classical gap horizontal}}{0.5cm}}}%
    }%
  }
    \anchor{dendtwo}{%
    \trianglepoints
    \pgf@process{\pgfpointintersectionoflines%
      {\pgfpointadd{\centerpoint}{\pgfmathrotatepointaround{\apexanchor}{\pgfpointorigin}{\rotate}}}%
      {\pgfpointadd{\centerpoint}{\pgfmathrotatepointaround{\lowerleftanchor}{\pgfpointorigin}{\rotate}}}%
      {\pgfpointadd{\centerpoint}{\pgfpoint{\pgfkeysvalueof{/tikz/commutative diagrams/classical gap horizontal}}{0cm}}}%
      {\pgfpointadd{\centerpoint}{\pgfpoint{\pgfkeysvalueof{/tikz/commutative diagrams/classical gap horizontal}}{0.5cm}}}%
    }%
  }
  \anchor{dendone}{%
    \trianglepoints
    \pgf@process{\pgfpointintersectionoflines%
      {\pgfpointadd{\centerpoint}{\pgfmathrotatepointaround{\apexanchor}{\pgfpointorigin}{\rotate}}}%
      {\pgfpointadd{\centerpoint}{\pgfmathrotatepointaround{\lowerleftanchor}{\pgfpointorigin}{\rotate}}}%
      {\pgfpointadd{\centerpoint}{\pgfpoint{-\pgfkeysvalueof{/tikz/commutative diagrams/classical gap horizontal}}{0cm}}}%
      {\pgfpointadd{\centerpoint}{\pgfpoint{-\pgfkeysvalueof{/tikz/commutative diagrams/classical gap horizontal}}{0.5cm}}}%
    }%
  }
}
\end{lstlisting}
\end{FullCode}
The \texttt{r} and \texttt{u} wires just go in the opposite direction to the \texttt{l} and \texttt{d} wires. So we don't need to calculate anything else, just equate them.
\begin{FullCode}
\begin{lstlisting}
%aliases for r and u wires
\pgfdeclareanchoralias{isosceles triangle}{dendtwo}{ustartwo}
\pgfdeclareanchoralias{isosceles triangle}{dendone}{ustartone}
\pgfdeclareanchoralias{isosceles triangle}{dstarttwo}{uendtwo}
\pgfdeclareanchoralias{isosceles triangle}{dstartone}{uendone}
\pgfdeclareanchoralias{isosceles triangle}{lendtwo}{rstartwo}
\pgfdeclareanchoralias{isosceles triangle}{lendone}{rstartone}
\pgfdeclareanchoralias{isosceles triangle}{lstarttwo}{rendtwo}
\pgfdeclareanchoralias{isosceles triangle}{lstartone}{rendone}
\makeatother
\end{lstlisting}
\end{FullCode}

We now have a shape that we can use as the base of a gate. It probably isn't quite styled how you might want it, so it's worth setting up a style.

\begin{FullCode}
\begin{lstlisting}
\tikzset{
  trimeter/.style={
      thickness, % gets the line width correct
      filling, %inherits filling. Typically background colour (default: white), but could be transparent.
      shape=isosceles triangle, %this is your shape name
      draw=black, %don't forget this, or you won't see much!
      inner sep=2pt
  }}
\end{lstlisting}
\end{FullCode}

Finally, define the code that will let you run the gate.
\begin{Code}[0.3]
\DeclareExpandableDocumentCommand{\mygate}{O{}{m}}{%
|[trimeter,#1]| {#2}}

\begin{quantikz}[classical gap vertical=0.05cm]
\lstick{\ket{0}} & \mygate{1} & \setwiretype{c} \\
&\gate{U}\wire[u]{c} &
\end{quantikz}
\end{Code}
Your function will work just like any of the single-qubit measurement commands: per-gate styling can be supplied via the optional argument. If your gate does not require the text label, you can remove the \texttt{\#2} from definition, but you must not remove the \texttt{\{m\}} from the argument specification.

It's possible to define a multi-qubit version in a very similar way, although they often end up looking a bit ugly, and non-straight edges are unlikely to fit correctly with incoming wires (which will be most noticeable if you activate transparency). It's why we don't officially have a multi-qubit version of \verb!\meterD!, although you may be able to get a good enough version working for your purposes.

\begin{FullCode}
\begin{lstlisting}
\DeclareExpandableDocumentCommand{\mygate}{O{}O{2em}O{1.5em}m}{%
\gate[#1,ps=trimeter,disable auto height][#2][#3]{#4}
}
\end{lstlisting}
\end{FullCode}

\section{Converting from QCircuit}

I've updated all of my existing teaching materials from QCircuit to the original Quantikz with very little trouble. There are a few standard replacements:
\begin{center}
\begin{tabular}{c|c}
QCircuit notation & Quantikz notation \\
\hline
\verb!\QCircuit @C=n @R=m {#}! & \verb!\begin{quantikz}[row sep=m,col sep=n]#\end{quantikz}!	\\
\verb!\multigate{n}! & \verb!\gate[n+1]!	\\
\verb!\targ! & \verb!\targ{}! \\
\verb!\control! & \verb!\control{}! \\
\verb!\meter! & \verb!\meter{}! \\
\verb!\measureD! & \verb!\meterD{}! \\
\end{tabular}
\end{center}
Updating the \verb!\gategroup! command requires a little more care since the first two arguments have to be removed, and the command placed in the correct cell, at which point \verb!\gategroup{i}{j}{k}{l}{m}! becomes 
\begin{center}
\verb!\gategroup[k+1-i,steps=l+1-j]!.
\end{center}
My primary use of gategroup was to achieve the effect now achieved with \verb!\lstick[k+1-i]!.

It should not be necessary to use \verb!\ghost! commands in the way they were used in QCircuit.

\subsection{Converting from Quantikz to Quantikz2}

Quantikz was originally written aiming to maintain compatibility with QCircuit. However, over time, it has proved desirable to add a number of new features. Some of these have necessitated breaking this backwards compatibility. This makes upgrading a little more fiddly in (probably) rare cases.

\begin{itemize}
  \item The original quantikz allowed you to use the tikzcd environment. This is no longer supported. You should always use the quantikz environment in place of the tikzcd environment.
\item Individual qw/cw commands are unnecessary. However, if there were places where you deliberately left a part of the circuit without a wire, that will need to be updated as everything has a qw by default.
\item Labels on e.g.\ \verb!\meter! are now in math mode where before they were in text mode. You'll get an error if your label now starts with \$.
\item \verb!\ctrlbundle! is obsolete. It will be automatically selected if appropriate. Just use \verb!\ctrl!.
\item \texttt{ampersand replacement} is unlikely to work (at least in the quantikz environment), but shouldn't be necessary.
\item Many of the styling options have been changed -- rather than inputting many of them directly like \verb![s]!, they can be accessed with a key \verb![style=s]!. This allows for much greater capacity for variation in the future. If you didn't use the options, then no changes are needed.
\item There are some instances where the original quantikz let you get away with a call such as \verb!\targ! instead of \verb!\targ{}!. Quantikz2 tends to be less forgiving.
  \end{itemize}



\section{Troubleshooting}\label{sec:trouble}
\begin{itemize}
\item Have you checked that all commands that need them are followed by an empty argument, \{\}? Things like \verb!meter!, \verb!\control! (basically, those that can accept an optional styling parameter) look like they don't take any parameters, but they have to be followed by the pair of braces or you'll get very odd effects.
\item If you get a whole bunch of unexpected text in one of your cells instead of a gate on an extremely wide gate, make sure that the gate command (or any other command indicated by \faToggleOn) is the first command in the cell, and that other commands (such as \verb!\qwbundle!) appear after.
\item If you're getting errors about cells not being found (and especially if you're doing any slicing, or gategroups), check that your last row doesn't end with \textbackslash\textbackslash, and make sure that your last row contains as many cells (even if they're empty) as there are columns in the circuit.
\item I am not aware of any specific compatibility problems with beamer, tabular, align etc.\ in the current version. 
\item If you're using transparency, and the width of gates seems to be greater than you expected, it may be worthwhile removing the .aux file and recompiling. If your tex editor isn't good at resetting the .aux file, the system may be remembering older widths.
\item Package load order: I've had reports that if you load certain packages in the wrong order it can create weird errors. For example, if you load the package cleveref after quantikz, and then use a split environment, it can lead to the error ``Only one \# is allowed per tab.''. Change the load order and it goes away. I have no idea why this happens.
\item If you are trying to submit to a Springer journal, they seem to actively prevent you from using tikz. You have to produce a separate image for each circuit. The externalisation options in tikz may be very helpful here to take your existing code and produce a pdf output of each circuit.
\end{itemize}

For any bug reports (please make sure you've checked the above list first!) or feature requests, please contact alastair.kay@rhul.ac.uk.

\section{Web-based Interface}\label{sec:web}

There is an experimental web-based interface available at \url{https://www.ma.rhul.ac.uk/akay/quantikz/}.

\begin{center}
\includegraphics[width=8cm]{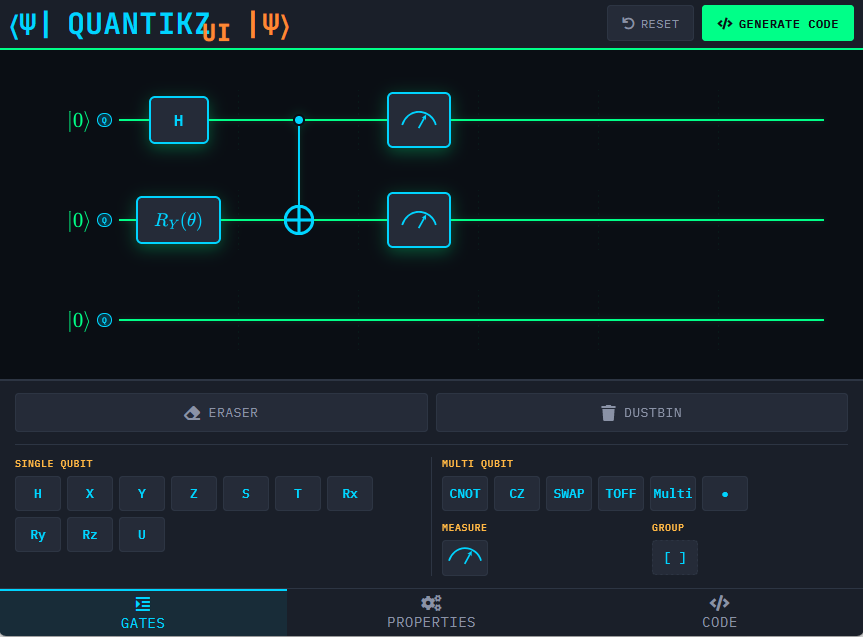}
\end{center}

This only provides a limited subset of quantikz' capabilities, but should be good for most uses (and if you want to go beyond that, you're probably ready to get stuck into the code!). The process should be fairly straightforward:
\begin{itemize}
  \item In ``Properties'', on the ``circuit'' tab, select the number of wires that your circuit uses.
  \item Wires default to being of the quantum type. If you want to change wire type, click the small pill at the start of each wire. It will cycle through the different wire types available (right-click cycles in the opposite direction).
  \item On the ``circuit'' tab, choose if you want to add slices.
  \item Each wire is given a default label of $\ket{0}$. Click on the label to change the text. You can also make it span multiple lines.
  \item If you want to add a label at the end of a wire, click on the label at the start of the wire and choose ``Add Rstick''. Then you can choose its label. If you want to add an end-label to a wire that is not the top wire when a left-label spans multiple qubits, there is a clickable area in about the right place.
  \item To add a single-qubit gate, simply drag and drop from the ``Gates'' palette. If it needs to go between two existing gates and there isn't a gap, if you hover at the midpoint, you'll see an orange bar appear to indicate that you will insert a new column. Click on the gate to edit its label.
  \item Measurement gates and multi-qubit gates work just the same.
  \item For a controlled-gate or swap gate, you drag to specify the point of the (first) control. Once you have dropped that, the mouse point changes to a cross-hair. Click where you want the next control (toffoli only) and then the target location.
  \item Gate group works similarly -- drag and drop the ``group'' element to where you want the top-left corner of the box to be. The mouse pointer changes to a cross-hair. Click where you want the bottom-right corner to be. Click on the text ``Group'' to change the text of the label.
    \begin{center}
\includegraphics[width=8cm]{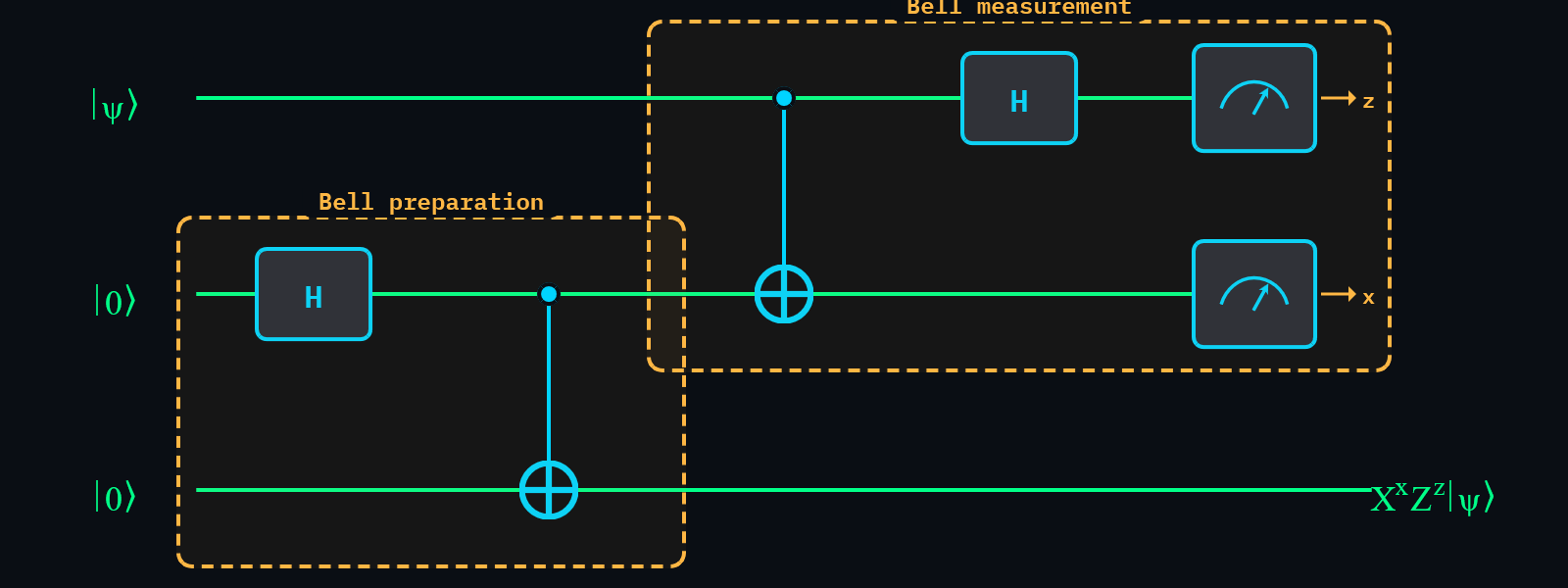}
\end{center}
  \item If you make any mistakes, you can move circuit elements around by dragging them (on a touch device, you may have to hover over the gate for a little while before it gets selected). You can delete them by dragging them onto the dustbin on the gates palette. Alternatively, use the eraser tool: click a gate to delete it. Click an empty wire and the next elements will shift one cell to the left (if possible).
  \item Once you're done, it's time to save your code! Click the ``Generate Code'' button and copy the provided code into your preferred text editor (don't forget to include the quantikz2 package in your document preamble). From the same window, you can export PNG, which will make an image of the circuit you've created. Either save it or copy it to the clipboard.
  \begin{center}
\includegraphics[width=8cm]{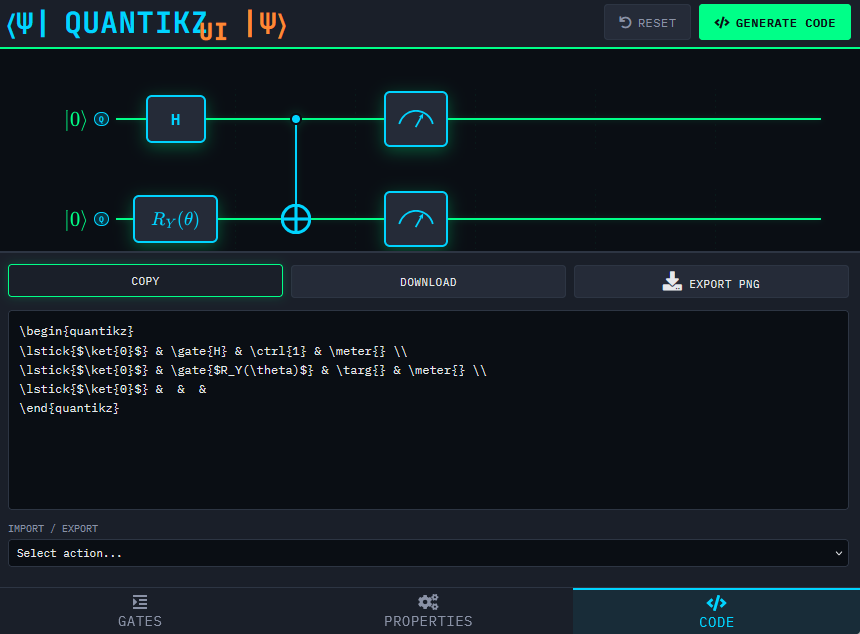}
\end{center}
\end{itemize}

Theoretically, it is also possible to import and export QASM and Qiskit circuits. This feature has not been extensively tested because I am not sufficiently familiar with these file formats.

\section{Citation}

If you found this package useful, please consider citing the arXiv version of this document, arXiv:1809.03842.

\end{document}